\documentclass[a4paper,fleqn,usenatbib]{mnras}

\usepackage[T1]{fontenc}
\usepackage{ae,aecompl}

\usepackage{graphicx}	
\usepackage{amsmath}	
\usepackage{amssymb}	

\title[Viscous Heating in V883 Ori]{Viscous heating as the dominant heat source inside the water snowline of V883 Ori}

\author[F. Alarc\'on et al.]{
Felipe Alarc\'on$^{1}$,
Sim\'on Casassus$^{2,3,4}$,
Wladimir Lyra$^{5}$,
Sebasti\'an P\'erez$^{4,6,7}$ and
\newauthor{Lucas Cieza$^{4,8}$}
\\
$^{1}$Department of Astronomy, University of Michigan,
323 West Hall, 1085 S University, Ave.,
Ann Arbor, MI 48109, USA \\
$^{2}$ Departamento de Astronom\'{\i}a, Universidad de Chile, Casilla 36-D, Santiago, Chile\\
$^{3}$ Data Observatory Foundation, Eliodoro Yáñez 2990, Providencia, Santiago, Chile\\
$^{4}$ Millennium Nucleus on Young Exoplanets and their Moons - YEMS, Chile\\
$^{5}$New Mexico State University, Department of Astronomy, PO Box 30001 MSC 4500, Las Cruces, NM 88001, USA\\
$^{6}$Departamento de Física, Universidad de Santiago de Chile, Av. Victor Jara 3659, Santiago, Chile\\
$^{7}$Center for Interdisciplinary Research in Astrophysics and Space Exploration (CIRAS), Universidad de Santiago de Chile, Chile\\
$^{8}$
Instituto de Estudios Astrofísicos, Facultad de Ingeniería y Ciencias, Universidad Diego Portales, Av. Ejercito 441, Santiago, Chile.}

\date{}

\pubyear{2022}

\begin{document}
\label{firstpage}
\pagerange{\pageref{firstpage}--\pageref{lastpage}}
\maketitle

\begin{abstract}
FU Orionis-type objects(FUors) are embedded protostars that undergo episodes of high accretion, potentially indicating a widespread but poorly understood phase in the formation of low-mass stars. Gaining a better understanding of the influence exerted by these outbursts on the evolution of the surrounding protoplanetary disc may hold significant implications for the process of planet formation and the evolution of disc chemistry. The heating due to outbursts of high accretion in FUors pushes the snowlines of key volatiles farther out in the disc, so they become easier to observe and study. Among the known FUors, V883 Ori is of particular interest. V883 Ori was the first FUor to show indirect evidence of a resolvable snowline beyond 40 au.  By introducing a radial-dependent model of this source including viscous heating, we show that active heating is needed to reproduce the  steep thermal profile of dust  in the inner disc of V883 Ori. Our disc modeling combines the effect of stellar irradiation and the influence on the disc shape caused by the outburst of accretion. The accuracy of our model is tested by comparing synthetic ALMA images with continuum observations of V883 Ori, showing that the model successfully reproduces the 1.3\,mm emission of V883 Ori at high spatial resolution. Our final predictions underline the importance of viscous heating as a predominant heat source for this type of object, changing the physical conditions (shape and temperature) of the disc, and influencing its evolution.

\end{abstract}


\section{Introduction}

FU Ori objects(FUors) are young stellar objects with episodes of
outbursts that occur when the accretion rates of the embedded  protostars are
enhanced by orders of magnitude \citep{HartRev2}. During these events, they reach
accretion rates up to $10^{-5}, 10^{-4}$ M$_{\odot}$ yr$^{-1}$
\citep{Hartrev1, Audard..2014, Fischer..PPVII}. In particular, accretion outbursts
also explain the low luminosity problem in the
stellar formation process of low-mass stars.  Assuming steady
accretion, the inferred accretion rates from the luminosity in
low-mass protostars are too low for them to reach their final masses
within the circumstellar disc lifetime \citep{LumProb, V346}. The
short episodes of high accretion rates solve this problem and
explain the low luminosities of the circumstellar discs during the low
accretion rate states. Nevertheless, understanding the stellar assembly is an ongoing effort \citep{Fischer..PPVII}.

Due to the outbursts of high accretion rate in FUors, we expect circumstellar discs to suffer significant changes in their physical conditions during this stage \citep{ARMANNREV, HartmannII}. \citet{YoungPSD} model circumstellar discs with radiative and episodic feedback, finding that episodic accretion changes the thermal and density profiles of the disc.  Moreover, \citet{NSAcc} analyze how young stellar objects accrete, and the changes induced on thermal profiles by different accretion rates and stellar masses.  Due to the high accretion rates reached by these objects, it is important to examine the relevant heat sources and how these sources shape the circumstellar discs and change their emission and evolution.

  Observations with the Atacama Larga Millimeter Array(ALMA) of V883 Ori, a FUor, have shown that V883 Ori has a very bright and optically thick core in the sub-mm continuum surrounded by an optically thin halo with an abrupt change of the spectral index at $\sim$40 au in millimeter wavelengths \citep{Snowline}. \citet{Snowline} associated the spectral index shift as an indirect tracer of the water snowline due to the sublimation of icy grains. Usually, the location of the water snowline for a passive disc of a T Tauri star is at a couple of au \citep{Lecar_SL}. However, V883 Ori's snowline beyond $\sim$40 au is an order of magnitude farther than the expected value, which cannot be explained by passive heating. i.e., only from stellar irradiation. Moreover, using molecular line emission, the snowline location in V883 Ori has been proposed to be even farther out at 80 au \citep{Tobin..Nature} and $\sim$100 au \citep{Merel, Leemker..2021}.

 Understanding the thermal structure of V883 Ori is essential for the interpretation of an increasing number of molecular line data on the object. The ongoing outburst in the V883 Ori disc has sublimated enormous amounts of volatiles typically trapped on ice mantles, rendering the source as a unique laboratory for the study of disc chemistry \citep{Merel,LEE_V883, Leemker..2021, Dary..et..al..2022, Tobin..Nature}.

There are multiple origins that can potentially explain the outward shift of the snowline in FUors. \citet{Har_Vis} used radiative transfer simulations to study the snowline location shifts when a flattened envelope is included in the disc. However, they found that an envelope would only shift the snowline by a couple of au. A shift of a few au is not large enough to explain the case of V883 Ori. Another explanation of the spectral index change can be linked to a more significant free-free emission in the inner disc due to higher temperatures instead of sublimation of icy grains \citep{mmcmFUORI}. Nonetheless, even in the presence of free-free emission, its main contribution in a circumstellar disc occurs at longer wavelengths than the ALMA band 6  observations ($\lambda = 1.3$ mm), making this scenario improbable for V883 Ori. Another possible reason for the snowline shift could be found in the accretion rate, which heats up the disc and raises the dust thermal emission. The inclusion of viscous dissipation can produce shifts of tens of au \citep{SL_MIN, Lecar_SL}, significantly boosting the brightness temperature and emission in the inner regions of actively accreting discs \citep{Labdon..21,Ueda..2023}. Therefore, we can explain the steep thermal profile in V883 Ori by including viscous heating in the disc.

The accretion outburst does not only change the temperature in FUors, it also changes the disc's physical structure. \citet{Bell_Flaring} show that viscous heating causes the flaring index to be variable in FUors, i.e., their aspect ratio has considerable radial variation, changing the incident angle of the stellar irradiation. For high accretion rate scenarios, the inner part of the disc follows a flared disc structure produced by the sharp radial gradient in viscous heating.  However, once the viscous heating stops being significant in the outer disc, the shape starts to change, producing a self-shadowing effect  \citep{BitschIII,BitschII}. Hence, we need to include the study of a non-constant flaring index structure in the modeling of this type of object. A variable flaring or shape in the disc can be key in explaining the brightness temperature and the steep dust continuum radial profile of V883 Ori. It will also have important effects on the chemical evolution of the disc by moving the water snowline \citep{Molyarova, Notsu..22}.

Here, we present the first model that reproduces the steep thermal profile for V883 Ori in millimeter dust continuum emission through the inclusion of active heating sources and heat diffusion in the disc which can play an important role in heat transport in discs \citep{Ziampras..2023}. Our model incorporates viscous heating as the main heat source in the inner disc, where the millimetric optical depth ($\lambda = 1.3$ mm) in V883 Ori has been estimated to be high enough to be emitting as a blackbody at radii up to $\sim$40 au.  Since the thermal structure and the shape are dependent on each other, we developed a radial-dependent algorithm that fits the millimeter dust continuum emission profile at each location. The algorithm fits the continuum intensity by modeling the physical temperature, its scale height, the dust distribution of the disc, and its accretion rate.

This paper is structured as follows. In Section  \ref{Sect2}, we describe a two-layer analytical model for the disc using viscous heating and we present the model that we used as input for the radiative transfer code \texttt{RADMC-3D} \citep{radmc3d}. In Section \ref{Sect3}, we describe the results of the radiative transfer model for the disc structure and the dust temperature together with the predictions for the millimeter emission. We discuss limitations in our model and further effects of viscous heating in Section \ref{Sect4}. Finally, we summarize our findings in Section \ref{Sect5}. 

\section{Methods}\label{Sect2}

\subsection{Two-layer Model Approach}\label{SubSect2.1}

To reproduce the dust emission in V883 Ori, our first modeling approach is done with a passive disc, i.e., only with irradiation heating. The model is resolved with an MCMC optimization of the parametric model from \citet{AESV}. The parametric model uses the self-similar solution of \citet{Lynden-Bell1974MNRAS} for the dust surface density:

\begin{equation}\label{eq:LB}
\Sigma = \Sigma_c\Big(\frac{r}{R_c} \Big)^{-\gamma} \exp\Big[-\Big(\frac{r}{R_c} \Big)^{2-\gamma} \Big],
\end{equation}

\noindent where $\Sigma_c$ is the dust surface density at the characteristic radius $R_c$. This first approach is done considering a constant flaring index $\psi$, i.e., the scale height of the disc is determined by the following relationship:

\begin{equation}
    h(r)=h_c \left(\frac{r}{R_c}\right)^\psi.
\end{equation}

\noindent We started with a standard flaring index $\psi = 1.15$ different from the fit from \citet{AESV} because a flaring index lower than one causes the disc to constantly self-shadow and cool down. This low flaring index is mostly produced by the overcompensation of the fit due to the lack of active heating sources in the disc.  A more detailed explanation for the parametric disc model and the results of the MCMC optimization can be found in \citet{AESV}.

 \begin{table}
 \caption{Parameters of Parametric disc in V883 Ori.}
\begin{center}
\begin{tabular}{ccc}
\hline
\hline
Parameter & Value  \\

\hline

Accretion Rate & $7.5\cdot10^{-5}$ M$_{\odot} \cdot$ yr$^{-1}$\\

$\psi$ & $1.15$\\ 

R$_c$, Characteristic Radius   & 31.6 au \\

h$_c$, Scale Height at R$_c$ & 4.17 au \\

M$_d$  & 0.6 M$_{\odot}$\\

$\gamma $ & 1.48 \\

\hline 

\end{tabular}
\end{center}
\label{tabla1}
\end{table}

 Before adding the viscous heating to the disc, we took the passive disc parameters from the MCMC optimization in Table \ref{tabla1} as the starting point of our model, finding that they were unable to recreate the centrally peaked and steep emission in the inner disc even with a variable aspect ratio. Thus, we add the viscous heating and let the accretion rate be variable with radius, so our model also predicts the vertical structure of the disc and the radial dependence of the accretion rate.  We estimate the radial parameters and associated uncertainties using a bootstrapping method sampling values from a distribution from a self-iterative method (more details in Appendix \hyperref[App]{A} ).

  V883 Ori is a young Class I source with a massive disc and a thin envelope, so the stellar size is quite uncertain. Nevertheless, we set its stellar size to $R_*=2.5\ R_{\odot}$ following the model from \citet{Gramajo..2014}. V883 Ori is part of the Northern region of the L1641 cloud cluster \citep{2008Reipurth}, which is close to the Orion Nebula Cluster (ONC). Thus, we adopted a distance of 388 pc to V883 Ori \citep{Kounkel..et..al..2017, Connelley_Reipurth}.
  
  The dynamical stellar mass from PV diagrams is consistent with $M_*=1.3\ M_{\odot}$ \citep{Snowline, Dary..et..al..2022}, while the starting disc mass  $M_d=0.6$ M$_{\odot}$ was obtained from the MCMC optimization made by \citet{AESV}. V883 Ori bolometric luminosity, $\sim 400$ L$_{\odot}$ \citep{1993ApJ_STROM}, is converted to a stellar accretion rate of $\dot{M}_* \sim 7 \times 10^{-5}$ M$_{\odot}$ yr$^{-1}$. \citet{Liu..2022} found a slightly larger accretion rate $\sim 1.1 \times 10^{-4}$ M$_{\odot}$ yr$^{-1}$ by modeling the observed energy spectrum. These observed accretion values are similar to the one inferred for FU Ori itself, which is $\sim \dot{M}_* \sim 4 \times 10^{-5}$ M$_{\odot}$ from a similar fitting procedure \citep{Perez..2020}. Despite the fact that it has been proposed that V883 Ori started its outburst of accretion over a century ago \citep{1993ApJ_STROM}, recent observations estimate its luminosity to be 212  L$_{\odot}$ \citep{Connelley_Reipurth}, which is probably showing the typical decay of FUors over time. Regardless of its luminosity decay, V883 Ori's luminosity is still very high for a 1.3 $M_{\odot}$ protostar, luminous enough to consider viscous heating as an extra heat source.

\subsubsection{Assumptions in our two-layer model}

We follow a two-layer model to reproduce V883 Ori's emission. One layer is the surface or photospheric layer, exposed to stellar irradiation; and the second layer is the midplane. We list the mathematical symbols used in the formulation of our model in Table \ref{tabla:symbols}. In our modeling, we assume the following conditions: 
\begin{enumerate}
    \item The disc is geometrically thin enough to be considered locally plane-parallel.
    \item The disc is axisymmetric.
    \item The disc is in local thermodynamic equilibrium and the viscous heating is distributed following the density vertical structure.
    \item We assume the radiation field to be isotropic, so we can use the Eddington approximation, i.e., the net radiation flux in the midplane is zero.
    \item For simplicity of the calculations we do not consider scattering in the analytical formulation.
\end{enumerate}

 \begin{table}
 \caption{List of symbols used in this work}
\begin{center}
\begin{tabular}{ccc}
\hline
\hline
Symbol & Definition & Description \\

\hline

$\Sigma$ &  & Surface Density  \\
$h$ &  & disc scale Height   \\
$\psi$ &  & Flaring Index   \\
R$_*$ &  & Stellar Radius   \\
$\dot{M}_*$ &   & Stellar Accretion Rate  \\
M$_d$ &  & disc Mass  \\
$\rho$ &  & Gas Density   \\
$\Sigma$ & =$\int \rho(r,z)dz$ & Surface Density   \\
$G$ &  & Gravitational Constant   \\
\hline
$T_{\rm irr}$ & Equation \ref{stel_plus_flar} & Irradiation ``Effective'' Temperature   \\
$T_{\rm acc}$ & Equation \ref{Vis_Heat} & Accretion ``Effective'' Temperature   \\
$T_{\rm bg}$ &  & Background Temperature   \\
$T_{\rm mid}$ & Equation \ref{eq: T_MID} & Midplane Temperature    \\
$q_{\rm irr}$ & Equation \ref{stel_plus_flar} & Irradiation Heating  \\
$q^+$ & Equation \ref{Vis_Heat} & Viscous Heating    \\
$\Gamma$ & Equation \ref{Vis_Final} &Viscous Energy Dissipation \\
$F_z$ & Equation \ref{eq:: T_mid}  & Vertical Midplane Heat Diffusion  \\
$F_r$ & Equation \ref{eq: F_r}  & Radial  Midplane Heat Diffusion  \\
\hline
$h_p$  & =4$h$ & Scale Height of Photosphere \\
$\Omega_K$ & =$\sqrt{GM_*/R^3}$ & Keplerian Rotation   \\
$\sigma_{\rm sb}$ &  & Stefan-Boltzmann's Constant  \\
$\nu$ &  & Viscosity  \\
\hline
$I_{\nu}$ &  & Specific intensity   \\
$J_{\nu}$ & Equation \ref{J_nu} & Mean Intensity   \\
$H_{\nu}$ & Equation \ref{H_nu}  & Eddington Flux  \\
$K_{\nu}$ & Equation \ref{K_nu} & K-integral, 2nd Moment   \\
$\kappa_{\rm rs}$  &   & Rosseland Mean Opacity \\
$\tau_{\rm mid}$ & =$\int_0^\infty \rho\kappa_{\rm rs} dz$ & Midplane Optical Depth   \\
\hline
$v_{\rm r}$ & Equation \ref{eq: v_r} & Radial Velocity    \\
$v_{\rm ff}$ & Equation \ref{eq: v_ff} & Free Fall Velocity    \\
\hline
\end{tabular}
\end{center}
\label{tabla:symbols}
\end{table}

\subsubsection{Heat Sources present in Active discs}

We consider three heat sources in our model. The first one is the irradiation heating that comes from the central star. The second heat source is the viscous heating produced by the accretion rate, and the third one is the background component coming from the interstellar radiation field. The last term is negligible compared to the other two heating sources. Nonetheless, we add it to set a constraint in the lower limit of the disc temperature.

The first heat source, irradiation heating, takes into account stellar irradiation and its dependence on the disc shape. Using the thin disc assumption and considering that the disc is being irradiated on both sides, the irradiation received by each side is approximately \citep{PPD_RT}:

\begin{equation}\label{stel_plus_flar}
q_{\rm irr} = \sigma_{\rm sb}T_{\rm irr}^4 = \sigma_{\rm sb}T_*^4\Big[ \frac{2}{3\pi^2}\Big(\frac{R_*}{r} \Big)^3\Big. +
\Big(\frac{R_*}{r} \Big)^2 \frac{h_p}{h}\Big(\frac{dh}{dr} -\frac{h}{r} \Big)\Big],
\end{equation}

\noindent where $r$ is the local radius from the star, $R_*$ the stellar radius, $h$ the local scale height and $h_p$ the photospheric scale height.

We introduce the variable $T_{\rm irr}$ in Eq.  \eqref{stel_plus_flar}; $T_{\rm irr}$ is the equivalent temperature for the incoming flux from the irradiation heating at one side of the disc. The first term in the right-hand side of the equation, $\frac{2}{3\pi}\Big(\frac{R_*}{r} \Big)^3$ (stellar term), is associated with the stellar size. The second term, $ \Big(\frac{R_*}{r} \Big)^2 \frac{h_p}{h}\Big(\frac{dh}{dr} -\frac{h}{r} \Big)$ (flaring term), is related to the flaring of the disc ($\frac{dh}{dr}$) and its local thickness ($\frac{h}{r}$), so it changes with the shape of the disc. In Equation \eqref{stel_plus_flar}, we add the factor $h_p$, or the height of the visible photosphere. The photospheric scale height was used by \citet{ChiangSED1997ApJ} to describe the location of the photosphere in circumstellar discs around protostars. In our model, the photospheric scale height is proportional to the scale height of the disc. At radii between 1 to 100 au, it is considered to be between  $4-5$ times the scale height of the disc ($h_p \approx 4-5 h$). In particular, our model uses $h_p = 4h$.  

 Besides the irradiation heating, we add the expression associated with viscous dissipation. The viscous heating for a standard $\alpha$-disc \citep{ShaSun} for each side of the disc in a ring of width $\delta r$ is \citep{Acc_Pow}:

\begin{equation}\label{Net_VIS_HEAT}
\delta Q^+ = \frac{9}{8}\Sigma\nu\Omega_K^22\pi r\delta r,
\end{equation}

\noindent where $\Sigma$ is the surface density, $\nu$ the viscosity, $r$ the radius and $\Omega_K = \sqrt{GM_*/r^3}$ the Keplerian angular velocity.

Taking the flux energy density of the viscous heating in Eq. \eqref{Net_VIS_HEAT}, we derive the expression:

\begin{equation}\label{Vis_Heat}
q^+ = \frac{9}{8}\Sigma\nu\Omega_K^2.
\end{equation}

Different radial and vertical prescriptions can be found depending on the disc structure assumed for the parameters $\Sigma$ and $\nu$. These parameters are difficult to constrain observationally, so we use the accretion rate instead, which also diminishes the number of free parameters in our model. To express the viscous heating as a function of $\dot{M}$, we use the equation relating the accretion rate with the viscosity and gas surface density for an accretion disc:

\begin{equation}\label{acc_visc}
\nu\Sigma = \frac{\dot{M}}{3\pi}.
\end{equation}

The viscous heating then becomes a function of the accretion rate and the Keplerian angular velocity. Combining Equations \eqref{Vis_Heat} and \eqref{acc_visc} the final equation for the viscous heating is  the following:  

\begin{equation}\label{Vis_Heat2}
q^+ = \frac{3}{8\pi}\dot{M}\Omega_K^2 = \sigma_{\rm sb}T_{\rm acc}^4.
\end{equation}

In Equation \eqref{Vis_Heat2}, $T_{\rm acc}$ represents the viscous heating as an effective accretional temperature. In this case, most of the viscous dissipation will be embedded inside the disc, below the photosphere. By using the diffusion approximation, the viscous heating will eventually reach the upper layers, heating them up. Given that it comes from the potential energy released by the gas being accreted, we assume that the vertical distribution of the viscous heating follows the density distribution.
We have to multiply our viscous heating expression in Equation \eqref{Vis_Heat2} by two to account for both disc halves. Therefore, the viscous energy dissipation per unit volume per unit mass in our model is:

\begin{equation}\label{Vis_Final}
\Gamma(r,z) = \frac{3}{8\pi}\dot{M}\Omega_K^2\frac{\rho(r,z)}{\Sigma(r)}, 
\end{equation}

\noindent where $\rho(r,z)$ is the density and $\Sigma(r)$ the surface density of the disc. The surface density in our model is a free parameter, but we expand it vertically assuming hydrostatic equilibrium. Hence, we assume a Gaussian distribution for the vertical structure of the disc using the scale height as the dispersion, i.e., $\rho(r,z) = \frac{\Sigma(r)}{h\sqrt{2\pi}}\exp(-z^2/2h^2)
$.

\begin{figure}
 \includegraphics[width=1\linewidth]{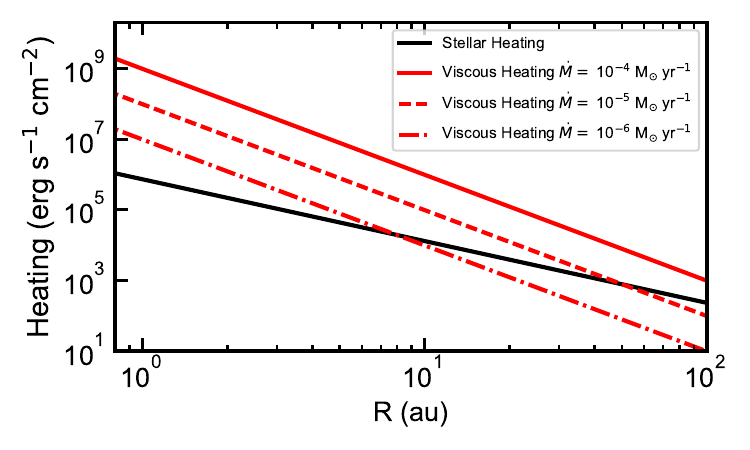}

\caption{Comparison between the stellar irradiation flux ($\sigma_{\mathrm{SB}}T^4_{\mathrm{irr}}$) and the viscous heating for three different constant accretion rates in a 7000 K star with 1.3 M$_{\odot}$. We observe that viscous heating dominates in the inner regions of  discs with strong accretion ($r<$10 au).}
\label{fig:V_vs_S}
\end{figure}

  As expected, the accretion rate is closely linked to viscous heating. Equations \eqref{Vis_Heat2} and \eqref{Vis_Final} show a linear dependence between the viscous heating and the accretion rate. If we consider a constant accretion rate, we would have a radial dependence of the form $q^+\propto \Omega_K^2 \propto r^{-3}$. Because of the cubic dependence of the viscous heating, its energy input will decay faster with radius than the stellar heating that follows the square-power law. Figure \ref{fig:V_vs_S} shows the comparison between stellar irradiation heating and viscous heating for a star with T$_* =$7000 K and R$_* = 2.5$R$_{\odot}$.  Under highly viscous conditions, accretion rates of FUors can reach values   $>10^{-4}$ M$_{\odot}$ yr$^{-1}$ \citep{NSAcc, Audard..2014}. When the accretion rate is of the order of $10^{-4}\ $M$_{\odot} \cdot$ yr$^{-1}$, the viscous heating overrules any other heat source to distances of the order of tens of au or even a hundred au. 
  
We also define a temperature set by the background radiation. This background radiation field sets a lower limit for the disc temperature, and is added as an external heat source  as follows:
  
\begin{equation}\label{eq: T_b}
 q_{\rm bg} = \sigma_{\rm sb}T_{\rm bg}^4,
\end{equation}

\noindent where we set $T_{\rm bg}=10$ K. 

Given the large thermal gradients produced by viscous heating and the high density of the disc, we included radial and vertical diffusion in the disc midplane as key energy transport mechanisms in addition to the previously mentioned heating sources. We describe the implementation of the radial and vertical diffusion in further detail in Sections \ref{Vertical} and \ref{Radial} respectively. 

\subsubsection{Effective Temperature}

We assume that the dust emission is thermal, i.e. $S_\nu= B_\nu $. Thus, we consider the effective temperature as the brightness temperature at millimetric wavelengths.  For the optically thick regions ($\tau\gtrsim$ 1) we use the brightness temperature as the dust temperature, while for the optically thin region, we correct the continuum attenuation by optical depth effects.

 The input energy at a given radius is the viscous heating from the disc's interior plus the irradiation heating from the star and what we have called the background radiation from the medium. The energy output at a certain location is just given by $T_{\rm eff}$:

\begin{equation}\label{eq: Energy_Eq}
q^+ + \sigma_{\rm sb}T_{\rm irr}^4 + \sigma_{\rm sb}T_{\rm bg}^4 = \sigma_{\rm sb}T_{\rm eff}^4.
\end{equation}

The left-hand side of Eq. \eqref{eq: Energy_Eq} is the input of energy in the disc and the right-hand side is the continuum emission.  From the energy conservation equation (Eq. \eqref{eq: Energy_Eq}) we obtain the relation for the radiative equilibrium of the disc.

For active discs, such as V883 Ori, viscous heating comes from the release of potential energy produced by the accretion rate. The fact that the dust emission is so steep and passive heating is not able to explain the brightness temperature means that $T_{\rm acc}^4 \gg T_{\rm bg}^4 + T_{\rm irr}^4 $. We assume that the brightness temperature($T_{\rm B}$) of the emission in the optically thick region should match the accretional temperature of the dust within a small margin, i.e., we enforce the condition that $T_{\rm eff} \approx T_{\rm acc}$ as our initial estimation of the local accretional temperature at each radius. So, we model the dust emitting as a blackbody at a temperature $T=T_{\rm acc}$, i.e.,
 
 \begin{equation}\label{eq: T_bright_T_acc}
 B_{\nu}(T_{\rm B}) \approx B_{\nu}(T_{\rm acc})(1-e^{-\tau}) \approx B_{\nu}(T_{\rm acc}).
 \end{equation}
 
 \noindent Then, we use the accretional temperature to solve the accretion rate using the viscous dissipation in Eq. \ref{Vis_Heat2}. Therefore, the accretion rate that we infer from the brightness temperature in the inner disc is: 

\begin{equation}\label{M_dot}
\dot{M} = \frac{8\pi\sigma_{\rm sb}T_{\rm acc}^4}{3\Omega_K^2}.
\end{equation}

 With that parameter, we estimate a local surface accretion rate necessary to achieve the dust emission once the effects of irradiation and background heating are considered simultaneously for all radii, including the disc shape dependence of the irradiation term.

After we obtain the viscous dissipation from the brightness temperature, we calculate the midplane temperature using radiative transfer equations with the assumptions previously detailed to calculate the vertical and radial diffusion. We then use the midplane temperature to study the disc shape and scale height by assuming hydrostatic equilibrium.

\subsubsection{Vertical Heat Diffusion}\label{Vertical}
As opposed to passive discs, the midplane temperature in accretion discs can reach temperatures $>$1000 K \citep{D'Alessio2005}. To calculate the effect of viscous heating in different layers of the disc, we also consider the optical depth as a driver of heat diffusion. By using the dust surface density together with the Rosseland mean opacity, $\kappa_{\rm rs}$, we calculate the optical depth at each radius of the disc. We consider the temperature-dependent opacity prescription from \citet{Semenov..2003} for the heat diffusion in the disc. Once the opacity and dust surface density are accounted for, the optical depth to the midplane is calculated using $\tau_{\rm mid} = \frac{\kappa \Sigma}{2}$.

For the optically thick inner regions, the disc is much hotter in the midplane (where the viscous heating has been input) than in the disc atmosphere, showing an expected thermal inversion \citep{Calvet..92, D'Alessio..98}. With such a vertical thermal gradient, the disc is heated up from the midplane to the surface instead of typical passively heated discs. 

To solve the vertical structure of the disc temperature we employ the radiative transfer equations for a gray atmosphere in the vertical axis, as in \citet{Hubeny1990ApJ}. Considering  $\mu=\cos(\theta)$, with $\theta$ the angle of the incident radiation, we use the standard intensity moments: $J_{\nu}$ as the mean intensity, $H_{\nu}$ as the Eddington  flux, and $K_{\nu}$ as the vertical projection of the radiation stress tensor or K-integral, i.e.,

\begin{equation}\label{J_nu}
J_{\nu} = \frac{1}{2}\int_{-1}^1 I_\nu d\mu,
\end{equation}

\begin{equation}\label{H_nu}
H_{\nu} = \frac{1}{2}\int_{-1}^1 \mu I_\nu d\mu,
\end{equation}

\noindent and

\begin{equation}\label{K_nu}
K_{\nu} = \frac{1}{2}\int_{-1}^1 \mu^2 I_\nu d\mu.
\end{equation}

 Using the plane-parallel approximation for a thin disc with no scattering term, the radiative transfer equations become:

\begin{equation}\label{eq:I_eq}
\mu \frac{dI_{\nu}}{dz} = \rho\Big(\frac{j_{\nu}}{4\pi} - \kappa_{\nu}I_{\nu}\Big),
\end{equation}

\begin{equation}\label{eq:zm_eq}
\frac{dH_{\nu}}{dz} = \rho\Big(\frac{j_{\nu}}{4\pi} - \kappa_{\nu}J_{\nu}\Big),
\end{equation}

\begin{equation}\label{eq:fi_eq}
\frac{dK_{\nu}}{dz} = -\rho\kappa_{\nu}H_{\nu},
\end{equation}

\noindent where $\rho$ is taken as the gas density, $\kappa_{\nu}$ the opacity, and $j_{\nu}$ the  dust emissivity. We apply the Eddington approximation (radiation field is isotropic on each hemisphere) to substitute $K_{\nu}$ by the mean intensity $J_{\nu}$ with an Eddington factor $f_K = K/J = 1/3$, i.e.,

\begin{equation}\label{eq: Edd_fac}
K_{\nu} = \frac{J_{\nu}}{3}.
\end{equation}

\noindent Then, we employ the energy conservation equation between matter and radiation for a given $(r,z)$ to solve the radiative transfer equation and get the Eddington flux:

\begin{equation}\label{eq: En_conser}
0 = \Gamma(r,z) - \rho(j - 4\pi \kappa_J J),
\end{equation}

 \noindent where $\Gamma(r,z)$ is the vertical distribution of the viscous heating given by Eq. \eqref{Vis_Final} and $\kappa_J=\int \kappa_\nu J_\nu d\nu/J$. We resolve the vertical structure in the disc by integrating Equation \eqref{eq:zm_eq} in frequency space combined with Equation \eqref{eq: En_conser}. The integrated-frequency version of Equation \eqref{eq:zm_eq} is the mean vertical energy flux, which is $\Gamma(r,z)/4\pi$.  Integrating Equation  \eqref{eq:zm_eq}, from the disc surface down to the midplane, leads us to the Eddington flux as a function of vertical depth in the disc:

\begin{eqnarray}\label{eq:zm_eq_t}
H(z)& =& H(0) + \int \frac{\Gamma(z)}{4\pi} dz \\
& =& H(0) + \frac{q_{\rm acc}^+}{4\pi \Sigma }\int _0^z \rho dz \label{eq:zm_eq_t3}\\
&= &H(0) + \frac{q_{\rm acc}^+}{4\pi \Sigma }\Big(\frac{\int_0^{\tau_z} d\tau}{\kappa_{\rm rs}}\Big) \label{eq:zm_eq_t4}.
\end{eqnarray}

We use the definition for optical depth in our vertical reference system from the surface to the midplane, 

\begin{equation}\label{eq: tau}
d\tau = \rho\kappa_{\rm rs}dz,      
\end{equation}

\noindent to solve Eq. \eqref{eq:zm_eq_t4} as a function of $\tau$ for simplicity. Before finding the vertical thermal structure of the disc we derive the expression for the Eddington flux at each optical depth:

\begin{eqnarray}\label{eq:zm_eq_t2}
H(\tau) &= &H(0) + \frac{q_{\rm acc}^+}{4\pi \Sigma \kappa_{\rm rs} }\tau_z \\
&= &H(0) + \frac{q_{\rm acc}^+\tau_z}{8\pi \tau_{\rm mid}},
\end{eqnarray}

\noindent where $H(0)$ is the Eddington flux at the surface, and $\tau_z$, is the optical depth measured from the disc surface to the midplane. $H(0)$ is given by the energy balance between the effective flux from the energy being diffused from the disc's interior from the viscous heating diffusion, F$_{\mathrm{in}}$; and the external energy input at the disc surface from the stellar irradiation and the background field, F$_{\mathrm{out}}$. Using Equation \ref{eq: Energy_Eq}, the expression for $H(0)$ is then: 

\begin{eqnarray}
H(0) &= &\frac{F_{\mathrm{out}} - F_{\mathrm{in}}}{4\pi}  \\
&= &\frac{\sigma_{\rm sb}T_{\rm eff}^4 - \sigma_{\rm sb}T_{\rm irr}^4- \sigma_{\rm sb}T_{\rm bg}^4}{4\pi}\\
&= &\frac{\sigma_{\rm sb}T_{\rm acc}^4}{4\pi}. \label{Edd_Flux}
\end{eqnarray}

\noindent Then, we use the optical depth definition in Equation \eqref{eq: tau} and the Eddington approximation between $K$ and J in Eq. \ref{eq: Edd_fac} to integrate Equation \eqref{eq:fi_eq} through all frequencies  to solve for J:

\begin{eqnarray} \label{eq: Aux}
\int_{J(0)}^{J(z_{\rm mid})} dJ & = & 3\int_0^{z_{\rm mid}} H\rho \kappa_{\rm rs} dz \\
&= & 3\int_0^{\tau_{\rm mid}} H(\tau) d\tau.
\end{eqnarray}

\noindent Then, expanding $H(\tau)$ from Eq. \eqref{eq:zm_eq_t2} in Equation \eqref{eq: Aux} we get the following relationship:

\begin{eqnarray}\label{eq:fi_eq_t}
\int dJ& =& 3\Big( H(0)\tau_{\rm mid} + \frac{q_{\rm acc}^+}{8\pi \tau_{\rm mid}} \int_{0}^{\tau_{\rm mid}} \tau d\tau \Big) \\ & =& 3\Big( H(0)\tau_{\rm mid} + \frac{q_{\rm acc}^+}{16\pi} \tau_{\rm mid} \Big)
.
\end{eqnarray}

Our mean intensity field $J$ at the midplane, which is directly related to the vertical diffusion $F_{z}$,  results in:

\begin{equation}\label{eq: Final_J}
J(z_{\rm mid}) = J(0) + 3H(0)\tau_{\rm mid} + \frac{3\sigma_{\rm sb}T_{\rm acc}^4 }{16\pi}\tau_{\rm mid} = \frac{F_{z}}{\pi}.
\end{equation}
To solve for the temperature at the midplane we apply the following boundary condition at the surface of the disc including external heating sources \citep{PPD_RT}. The boundary condition assumes a Lambertian reflective surface, i.e. the reflection is isotropic, and it takes into consideration the irradiation effects on the disc. \footnote{A more thorough discussion about the boundary condition choice can be found in \citet[Section 3.1]{PPD_RT}}: 

\begin{equation}\label{Eq: J_0}
 J(0) = 2H(0) + \frac{\sigma_{\rm sb} T_{\rm irr}^4}{\pi} + \frac{\sigma_{\rm sb}T_{\rm bg}^4}{\pi}.
\end{equation}

\noindent The final expression for the vertical diffusion flux given our assumptions is:

\begin{equation}\label{eq:: T_mid}
F_{z} = \sigma_{\rm sb}\Bigg(\frac{3}{4}T_{\rm acc}^4\Big(\tau_{\rm mid} + \frac{2}{3} \Big) + T_{\rm irr}^4 + T_{\rm bg}^4 + \frac{3}{16}T_{\rm acc}^4 \tau_{\rm mid}\Bigg).
\end{equation}

 To calculate the optical depth in the millimeter regime of the inner disc, we use a mean opacity value without scattering $\kappa_{\rm mm}=0.02$ cm$^2\cdot$  gr$^{-1}$ \citep{Beckwith..et..al..1990}. This opacity value is used to calculate the physical temperature form the observed brightness temperature.

 \subsubsection{Radial Diffusion}\label{Radial}

  In the thick midplane, heat diffusion has a radial and a vertical component, any azimuthal component would be zero given our assumption of axisymmetry. We still have radial heat diffusion as an additional component of our model. Following the derivation in \citet{Casassus..et..al..2019}, the radial heat diffusion at a given radius $r$ is:

\begin{eqnarray}
    F_{\mathrm{r}}(r)& = &\frac{1}{r}\frac{\partial}{\partial r}r 2\pi h^2 \frac{16\sigma_{\rm sb} T^3}{3\kappa_{\rm rs}\Sigma}\frac{\partial T}{\partial r} \\
    &=& \frac{1}{r}\frac{\partial}{\partial r}r\sqrt{2\pi} h\frac{16\sigma_{\rm sb} T^3}{3\kappa_{\rm rs}\rho_0}\frac{\partial T}{\partial r}, \label{eq: F_r}
\end{eqnarray}

\noindent where $\rho_0$ is the density in the midplane, $\kappa_{\rm rs}$ the Rosseland opacity, $\frac{dT}{r}$ the radial thermal gradient at a given radius $r$, $T$ the midplane temperature as a function of $r$, and $h$ the local scale height. Radial diffusion attempts to smooth strong thermal gradients and discontinuities in the disc that can be caused by other effects, such as abrupt changes in dust opacity or physical star-disc interactions.

\subsubsection{Midplane temperature in the optically thick regions}

Once we have all the heating and diffusion terms, we calculate the midplane temperature considering both, the vertical term from Equation \ref{eq:: T_mid} and the radial diffusion term from Equation \ref{eq: F_r} balancing the energy input and output in the disc midplane. Thus, the midplane temperature, $T_{\mathrm{mid}}$, is:

\begin{equation}
    \sigma_{\mathrm{sb}}T_{\mathrm{mid}}^4 = F_z + F_r.
    \label{eq: T_MID}
\end{equation}

  Once $T_{\mathrm{mid}}$ is calculated, the local scale height is updated accordingly, which will change the calculated values for the aspect ratio,  $\frac{h}{r}$, and the flaring index. We iterate the procedure in order to get self-consistent results until $\vert T_{\mathrm{model}} - T_{\mathrm{brightness}} \vert <$0.1.

\subsubsection{Outer disc, the Optically Thin Regime}

In the outer disc, the measured brightness temperature is lower than the temperature of an irradiated disc, confirming that the disc has to be optically thin in this region ($\tau_{\rm mm}\lesssim 1$).  Therefore, we model dust emission differently. Given that the disc is more diffuse at this region, we apply a simplifying approach assuming that the disc is vertically isothermal, i.e., the midplane and surface temperature are the same.

We measure the optical depth in the outer disc by solving the radiative transfer equation $I_\nu = B_\nu(1-e^{-\tau_{\rm \nu}})$, from the previous fitting of V883 Ori's physical structure in the passively heated region. Therefore, the expression for the millimeter optical depth, $\tau_{\rm mm}$, is: 

\begin{equation}\label{eq: Opt_depth}
\tau = -\ln \Big(1-\frac{B_\nu(T_{\rm B})}{B_\nu (T_{\rm dust})}\Big).
\end{equation}

\subsection{Radiative Transfer Simulations}

We run radiative transfer simulations using the derived two-layer model including active heating to get the thermal structure of V883 Ori. We simulate the radiative transfer using \texttt{RADMC-3D} \citep{radmc3d} adding an additional heat source in the disc, as it was implemented by \citet{Hord_2017ApJ}. This approach includes the amount of energy added by the external heat source at every cell of the grid.  We include the extra heat source using the expression in Equation \eqref{Vis_Heat2} and the inferred accretion rate from the two-layer model fitting. We then expand it vertically as a 3D flux density using  Equation \eqref{Vis_Final}. Despite the fact that the vertical distribution is not well known, changing its vertical distribution between a uniform and a gaussian did not significantly change the synthetic millimeter continuum emission.

The numerical simulations are run in a numerical grid of $500\times256\times1$ cells in $(r,\theta,\phi)$ respectively. The radial cells follow a logarithmic sample ranging from 0.8 au to 180 au. The colatitudes values are linearly sampled from $0.1$ to $\frac{\pi}{2}$ assuming mirror symmetry to the other side of the midplane. The input density field of the radiative transfer simulations is the one obtained from the two-layer model. The gas and dust are vertically distributed with a gaussian centered at $z=0$ and the scale height as the dispersion. We use a variable flaring including the shadowing effect obtained from the self-consistent fit.

We calculate the thermal equilibrium using $10^{8}$ photon walkers for each run. Dust opacity is modeled using two different dust compositions. The first dust composition corresponds to dry grains with mass fractions of $50\%$ organics, $41\%$ astrosilicates, and $\sim8.8\%$ troilite. The second composition, the icy grains, has a water mass fraction of $20\%$, $40\%$ in organics, $33\%$ in astrosilicates, and $\sim7\%$ in troilite.  The opacity values were taken from \citet{Henning..Stognienko..1996} for the troilite and organics composition, from \citet{Draine..2003} for the astrosilicates, and \citet{Warren..Brandt..2008} for water ice. We set the radiative transfer with three iterations to update the dust populations for self-convergence. Wherever the dust temperature is lower than 150 K, the dust population is composed of icy grains, and if the temperature is higher than 150 K, it is composed of dry grains.

The dust size distribution follows a standard power law proportional to the dust grain size:

\begin{equation}\label{eq: dustsize}
n_0(a)dn \propto a^{-3.5}da,
\end{equation}

\noindent with $a$ the size of the dust particle \citep{Mathis..et..al..1977}. The dust grain sizes range from $a_{\rm min}= 0.05$ $\mu$ m to $a_{\rm max} = 2.5 $ mm for the icy large grains, while $a_{\rm max} = 0.5 $ mm for the dry grains inside the snowline, simulating an enhanced dust growth due to the water ice stickiness in the outer disc. 

Once the thermal equilibrium is calculated, we generate a synthetic continuum image for $\lambda=1.3$ mm. After that, we compare it with ALMA band 6 observations of V883 Ori by convolving the resulting image with the same elliptical beam of the observations. The elliptical beam of the observations has a major axis of $\approx$ 37 mas, a minor axis of $\approx$ 27 mas, and a position angle of 52 degrees.

\section{Results}\label{Sect3}
\begin{figure*}
 \includegraphics[width=1.\linewidth]{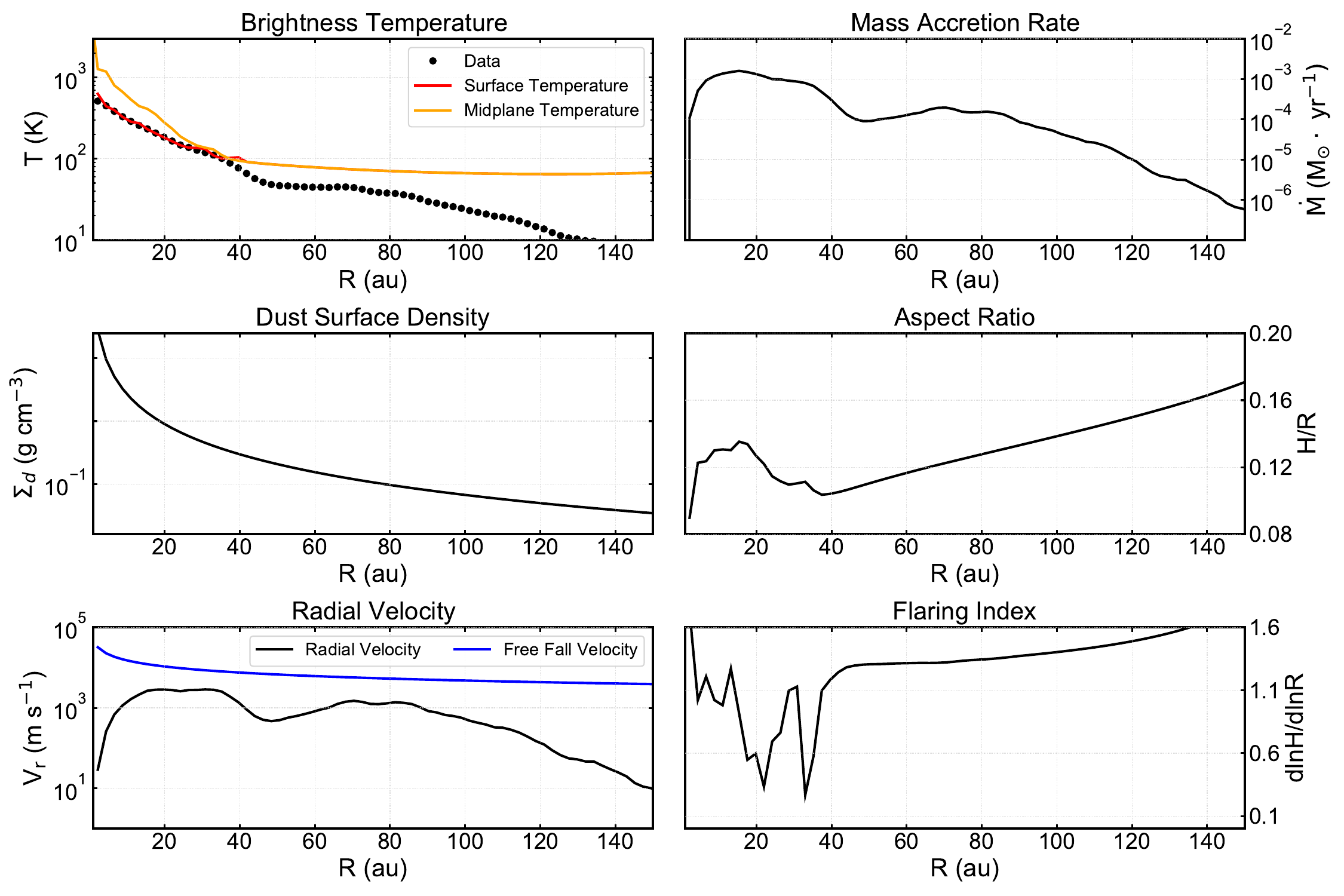}
\caption{Physical conditions of our model. The midplane temperature looks hotter than the surface temperature, as expected given the optical depth effects. The flaring index and the aspect ratio show the disc self-shadowing and thickening generated by viscous heating. The flaring index shows a wiggle behavior caused by abrupt changes in the temperature-dependent opacity. After 50 au, the viscous heating is negligible so the disc behaves as a passive disc and the accretion becomes uncertain at that point. We also include a comparison between the measured in-fall velocity and the gas free-fall velocity to verify that gas and dust are coupled so dust temperature is representative of the viscous dissipation.}
\label{fig: Self-Consistent}
\end{figure*}

\subsection{Radial Dependence of Flaring and Aspect Ratio}

 The solution in Figure \ref{fig: Self-Consistent} shows the self-shadowed regions of the disc and its variable flaring. There is a decline in the aspect ratio beyond the snowline where the accretion rate also starts to decrease. The change in the disc morphology is mainly caused by the decrease in the effect of viscous heating and the change in the opacity of the dust grains. As the disc gets hotter, it also gets geometrically thicker so the flaring index actually decreases when the viscous heating starts to decay. When the flaring index is less than one, the disc does not receive direct stellar irradiation. Thus, behind the hot and thick region heated by viscous heating, the disc gets colder due to self-shadowing, having additional strong implications for the photochemical reactions in these regions. in

 The flaring index in Figure \ref{fig: Self-Consistent} shows abrupt changes. Those jumps can be explained by the significant opacity changes which are temperature-dependent \citep{Semenov..2003}. One of those changes is strongly related to the water snowline. As water sublimates from icy grains, the opacity suffers drastic changes which translates into an abrupt jump in the disc shape as well, which is illustrated by the flaring index. 
 
 In the self-shadowed region, Figure \ref{fig: Self-Consistent} shows a thick disc between 10 and 30 au. This thick region causes the outward zones to be shadowed by stellar radiation. Additionally, our model predicts a wiggle in the accretion rate. The accretion rate reaches its peak in the plateau and starts an abrupt decay afterward, right at 40 au. The change in the disc's accretion rate could mean that there is an accumulation of material at the location of the spectral index shift. Such an effect could be key to understanding the variability of FUors and the effect of snowlines on dust evolution and planet formation. Moreover, shadowing regions produce variability through thermal-wave instability and radial evolution in the location of the snowlines \citep{okuzumi..22}.
 
\begin{figure}
 \includegraphics[width=1.\linewidth]{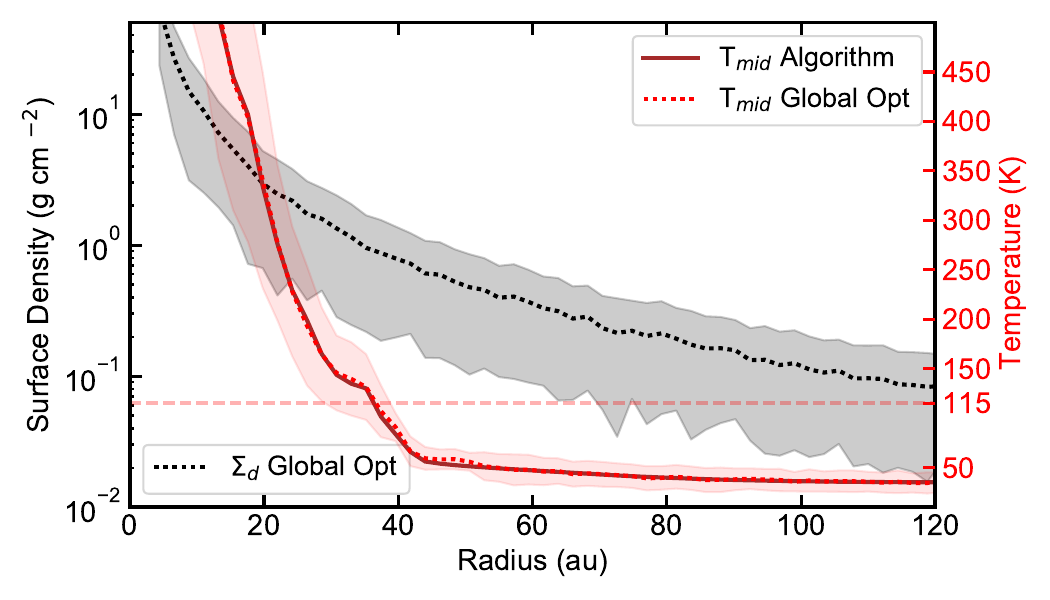}
\caption{Midplane temperature and dust surface density as a function of radius in V883 Ori. We show a comparison between the distributions obtained from the bootstrapping algorithm and the self-iterative method, which are in agreement with each other.}
\label{fig:bootstrapping}
\end{figure}

\subsection{Effects of Viscous Heating on Thermal Structure}

The first consequence of the addition of viscous heating is the increase of the dust temperature to values $\sim 1000$ K close to the star (see Figure \ref{fig: Self-Consistent} ). The high gas density and the uniform vertical distribution of viscous heating in the disc produce an almost constant  thermal structure at different layers of the disc. 

The results of the two-layer fit in Figure \ref{fig: Self-Consistent} show that the viscous heating is substantially more important at the inner 40 au. When the viscous heating is included in V883 Ori, the midplane becomes hotter than the surface layers. We also observe that the viscous heating increases the temperature at the disc midplane as far as 40 au from the star. For the outer part of the disc ($r>70$ au), the viscous heating stops being significant and the disc is almost isothermal in the vertical direction, so our initial assumption for the outer disc seems reasonable.

Figure \ref{fig:bootstrapping} shows the results of the fitting employing the bootstrapping method (more details in Appendix \ref{App:2}) with the respective uncertainties. Overall, the bootstrapping method matches the fit from the self-consistent approach, with a very good constraint of the dust temperature in
the midplane. We also notice that even though the surface density matches the assumed input, the uncertainties are larger, and therefore the dust distribution is less constrained in the optically thick region of the disc.

\begin{figure}
 \includegraphics[width=1.\linewidth]{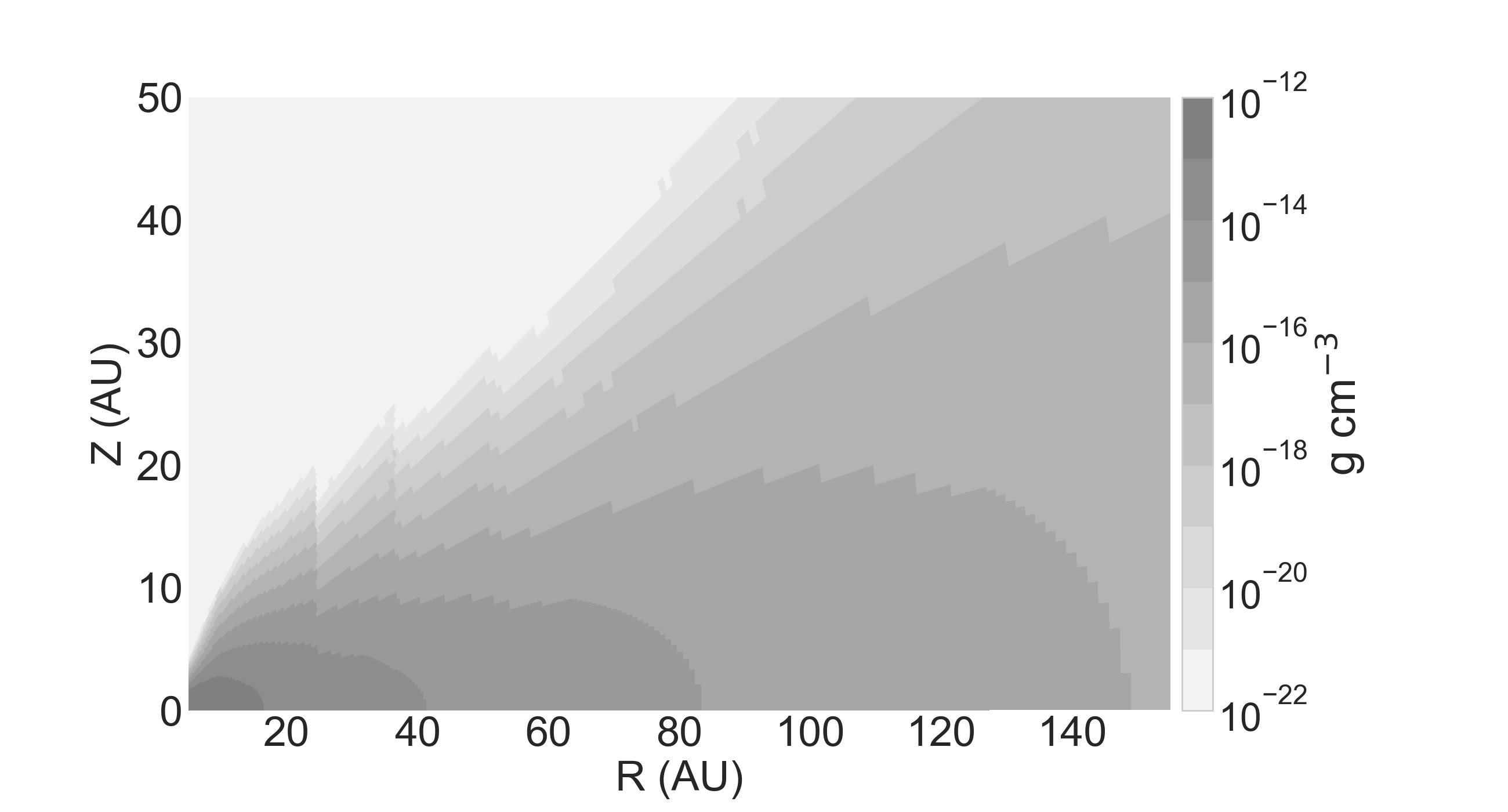}

\caption{Dust density distribution in our simulations including viscous heating. It illustrates the variable disc shape and the self-shielded regions caused by including viscous heating.}
\label{fig:Densities}
\end{figure}

\begin{figure*}
  \includegraphics[width=.48\linewidth]{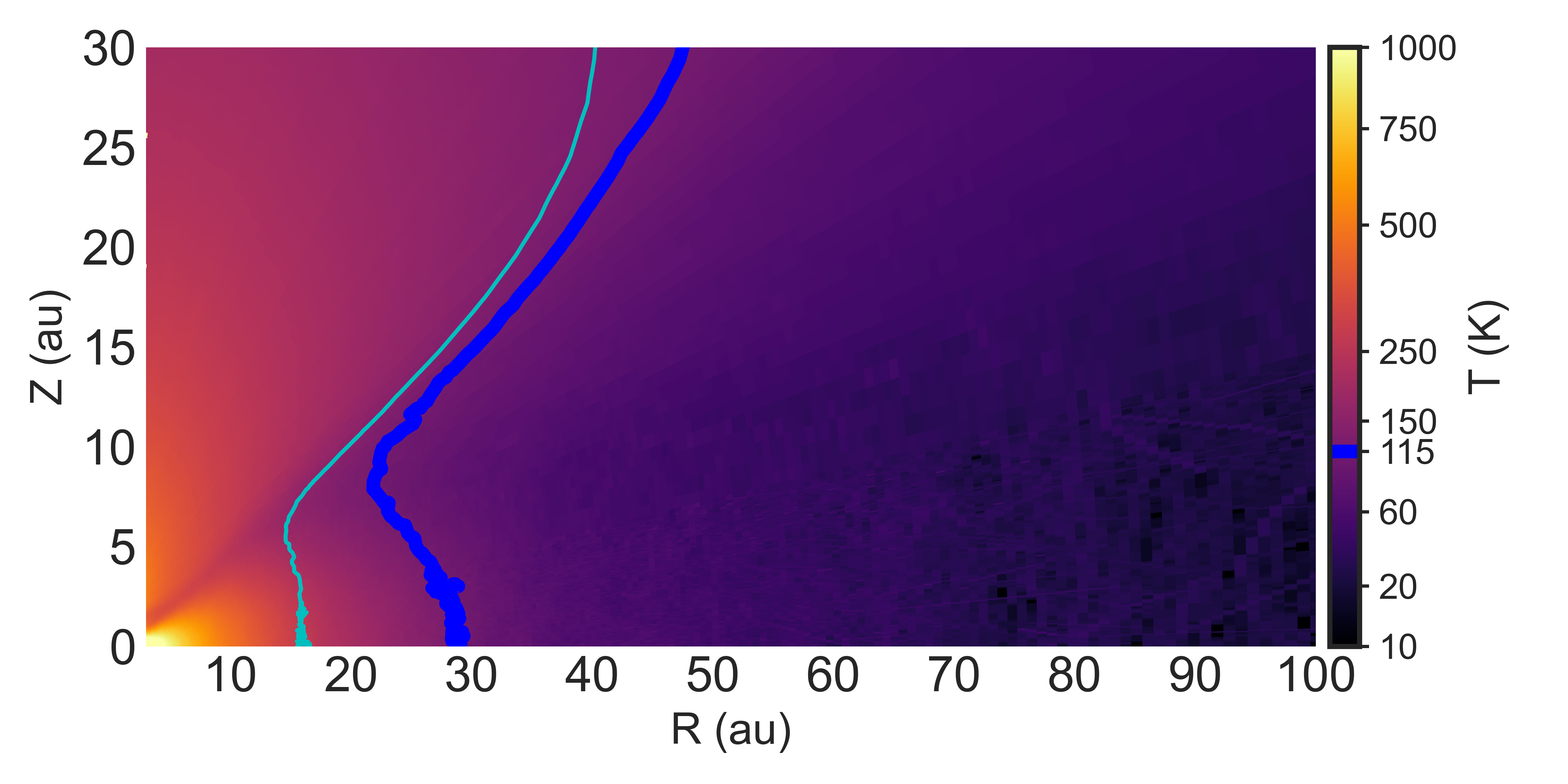}
  \includegraphics[width=.48\linewidth]{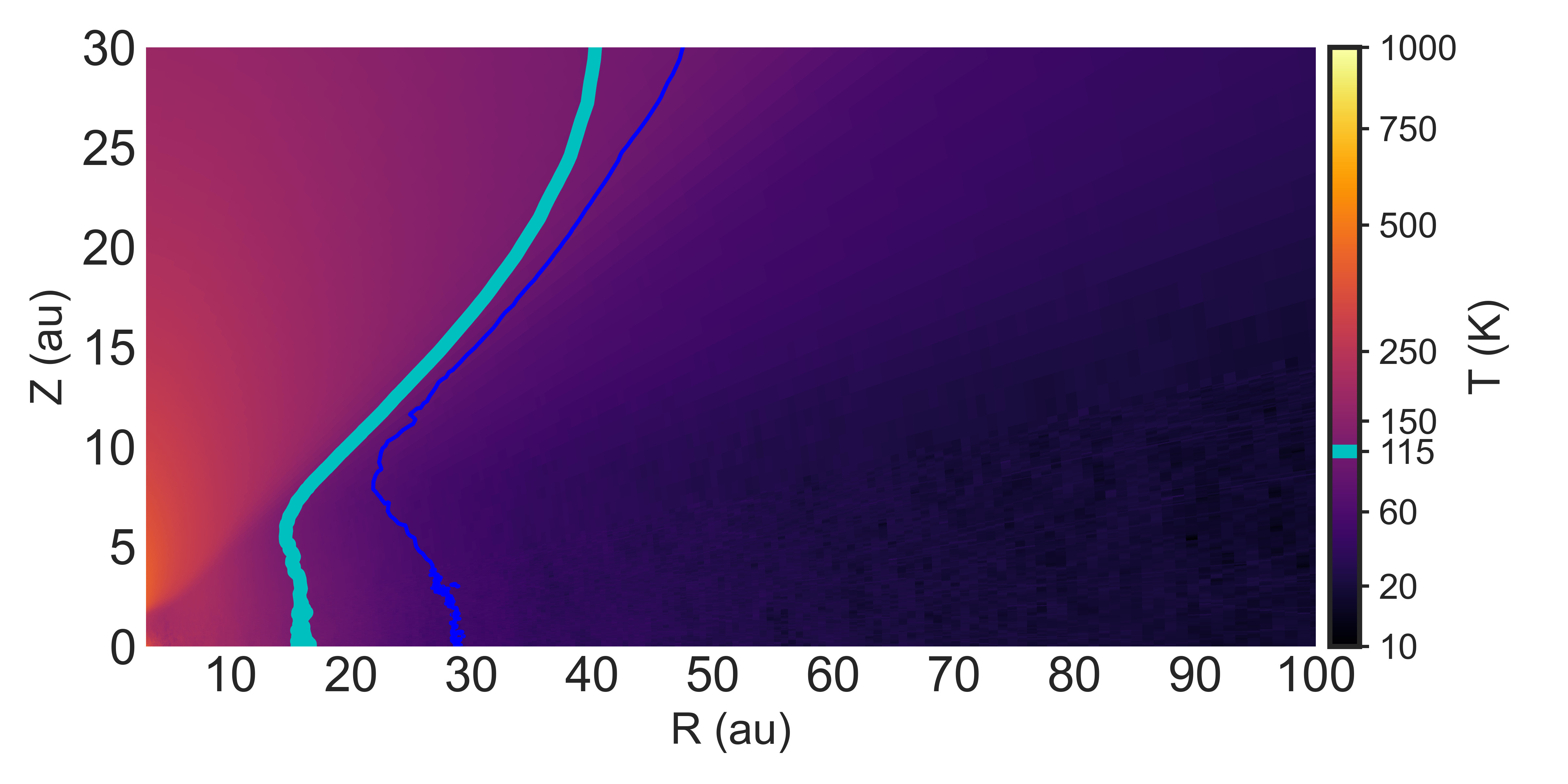}
  \caption{Dust temperature in our models turning on and off the
    viscous heating. The blue line is an isothermal contour at
    $T=115\,$K in the model with viscous heating added and the cyan the same isothermal contour in the model without viscous heating. We use 115 K as  the temperature at which \citet{Snowline} locate the spectral index shift in
    V883 Ori. \textbf{Left:} Image with viscous
    heating. \textbf{Right}: Image without viscous heating. We observe
    in the images that when we add viscous heating the snowlines get
    shifted by almost 20\,au.}
\label{fig:Temperatures}
\end{figure*}

We illustrate the input dust density meridional structure in Figure \ref{fig:Densities} showing the radially variable structure of the dust distribution in the V883 Ori. From comparing the effects of viscous heating in the disc temperature in Figure \ref{fig:Temperatures}, we observe that the 115 K layer is shifted by 20-30 au, almost matching the spectral index change in the disc. The role of viscous heating in pushing the condensation fronts outwards is proved by the radiative transfer models. Moreover, it is observed that the viscous heating affects the disc mostly in the inner regions($<$40 au), while the outer regions remain without significant thermal changes. At those radii, the irradiation heating dominates and the typical vertical structure of passive discs is recovered.

The simulations without viscous heating produce a dust temperature lower than the brightness temperature at almost every layer. Since the brightness temperature is a lower bound for the dust temperature, the conclusion of an extra heat source for the best fit of the data is reinforced. Irradiation heating by itself is not enough to explain the spiky radial continuum profile or the high brightness temperatures in the inner regions of V883 Ori (r$\lesssim$ 40 au).

\begin{figure}
 \includegraphics[width=1.\linewidth]{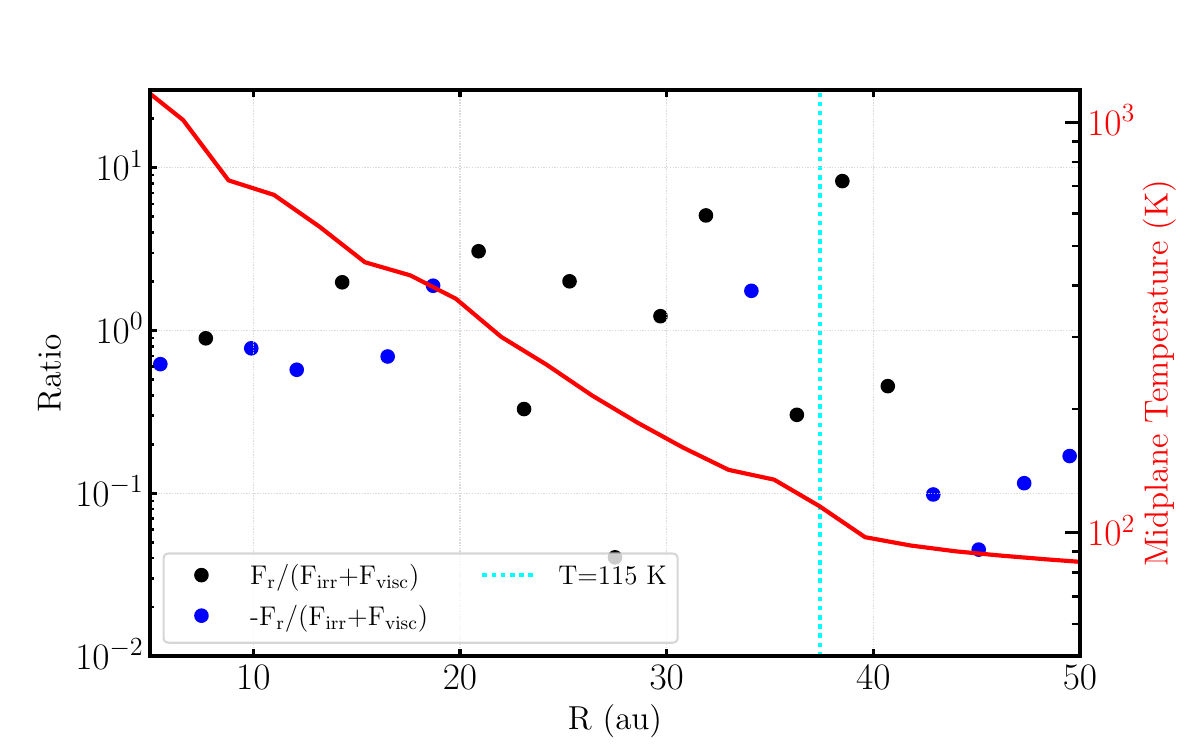}
\caption{Ratio between the radial flux inside and the heating terms considered in this work. For the self-shadowed region, the radial heat diffusion will be comparable to the heating terms so it is expected that it will raise the disc temperature. Radial heat diffusion seems to be important right outside the snowline as well. However, our opacity values change drastically in and out of the snowline due to the ice sublimation, causing the radial heat diffusion to be more significant.}
\label{fig: fr/fz}
\end{figure}

Due to the high midplane temperature caused by viscous heating and the high surface density of V883 Ori, we compare the different heating terms in order to gauge the relevance of radial diffusion in our model. Figure \ref{fig: fr/fz} shows the comparison between the radial heat diffusion (see Eq. \eqref{eq: F_r}) and the other heating terms, i.e., the viscous heating, $q^+$, and the irradiation heating, $\sigma_{\mathrm{SB}}T^4_{\mathrm{irr}}$. The Figure proves the fact that radial diffusion is comparable to vertical diffusion when the radial thermal profile of the disc is steep, attempting to counteract radial temperature differences and smoothing the radial thermal profile in both directions. The radial heat diffusion smooths strong thermal gradients that can be caused by obscured regions by self-shadowing effects. Nevertheless, as the disc thermal gradient flattens due to viscous heating decay and the disc starts to become optically thin, the radial diffusion flux becomes negligible.

\subsection{Accretion Rate as a Radial Function}

We show the local accretion rate in the disc as a function of radius in Figure \ref{fig: Self-Consistent}. We observe that the accretion rate reaches a peak value  $\sim 10^{-3}$ M$_{\odot}\cdot$ yr$^{-1}$  at the location of the plateau ($30<r<40$ au). Beyond the plateau, the accretion rate and the viscous heating are less strong, and the disc is passively heated. The fact that the accretion rate is not continuous throughout the disc means that there must be a pile-up of material. However, given that V883 Ori is a FUor, we expect that it has gone through a sequence of accretion outbursts, explaining the differences in the local accretion rate. 

As a safe check, we also measure the radial velocity from the accretion rate, 
\begin{equation}\label{eq: v_r}
v_r = \frac{\dot{M}}{2\pi\Sigma r},
\end{equation}

\noindent which is shown in Figure \ref{fig: Self-Consistent}. Our measured radial velocity is below the free fall velocity,

\begin{equation}\label{eq: v_ff}
v_{\rm ff} = \sqrt{\frac{2GM_*}{r}},
\end{equation}

\noindent even at the location where the peak of accretion occurs. Therefore, the gas motion is not related to virialization and the assumption of dust-gas coupling seems reasonable. So, the assumption of using the continuum emission as a tracer of viscous heating is reliable.

We calculate the accretion luminosity produced by viscous heating in the inner disc as another one of our safety checks. This accretion luminosity should not be higher than the measured bolometric luminosity for V883 Ori ,$L_*\sim 400$ L$_{\odot}$ \citep{Sandel..et..al..2001}. To calculate the accretion luminosity in the disc, we integrate the viscous heating  through concentric rings at each radius, i.e.,

\begin{equation}\label{eq: Acc_L}
L(r_1,r_2) = \int_{r_1}^{r_2}2 \cdot 2\pi r \frac{3}{8\pi}\dot{M}(r)\Omega_K^2(r) dr.
\end{equation}

\noindent The integrated accretional luminosity from Equation \ref{eq: Acc_L} is L$_{\rm acc}\approx 80$ L$_{\odot}$. Even if we consider that roughly half of the accretion luminosity should be emitted very close to the star inside the dust cavity, our integrated luminosity is less than $20\%$ of V883 Ori's bolometric luminosity. Thus, it is possible then that we are underestimating the contribution of other emitting components in V883 Ori that are not probed by millimeter wavelengths. Another probability is that the accretion luminosity in the inner disc is highly variable, which might explain why different measurements taken at different epochs give different values (e.g. L$_{\rm bol}$= 200 vs 400 L$_{\odot}$).


\subsection{Synthetic Predictions for V883 Ori}

\begin{figure}
 \includegraphics[width=1.\linewidth]{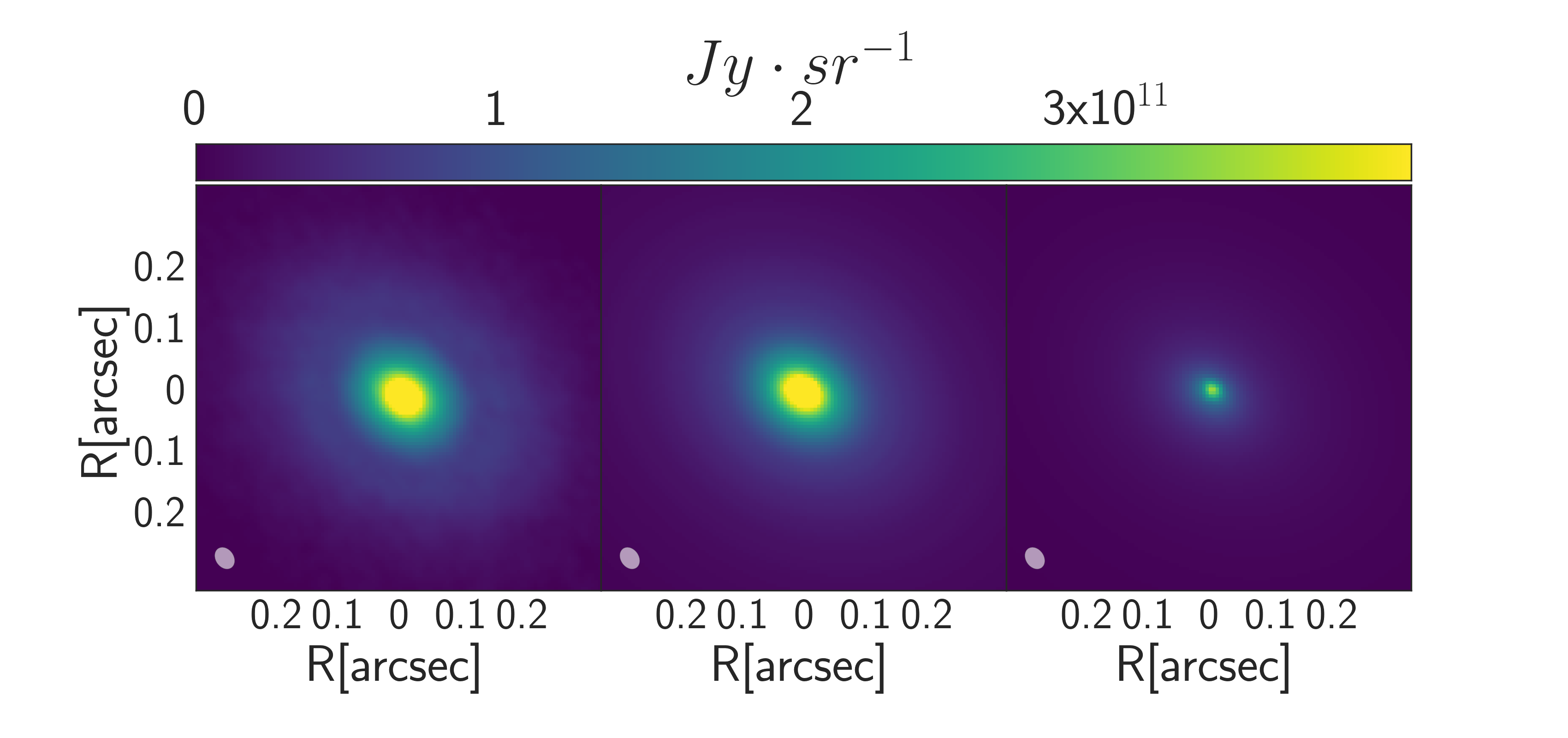}

\caption{Comparison of dust continuum images at $\lambda=1.3$ mm between the observations and our models with and without viscous heating. \textbf{Left}: Observations. \textbf{Center}: Image with viscous heating.  \textbf{Right}: Image without viscous heating. The images show that the addition of viscous heating matches the central peak emission of the observations, whereas the image without viscous heating is not able to match the central emission in the inner au of the disc.}
\label{fig: Images}
\end{figure}

\begin{figure}
\begin{center}
 \includegraphics[width=1.1\linewidth]{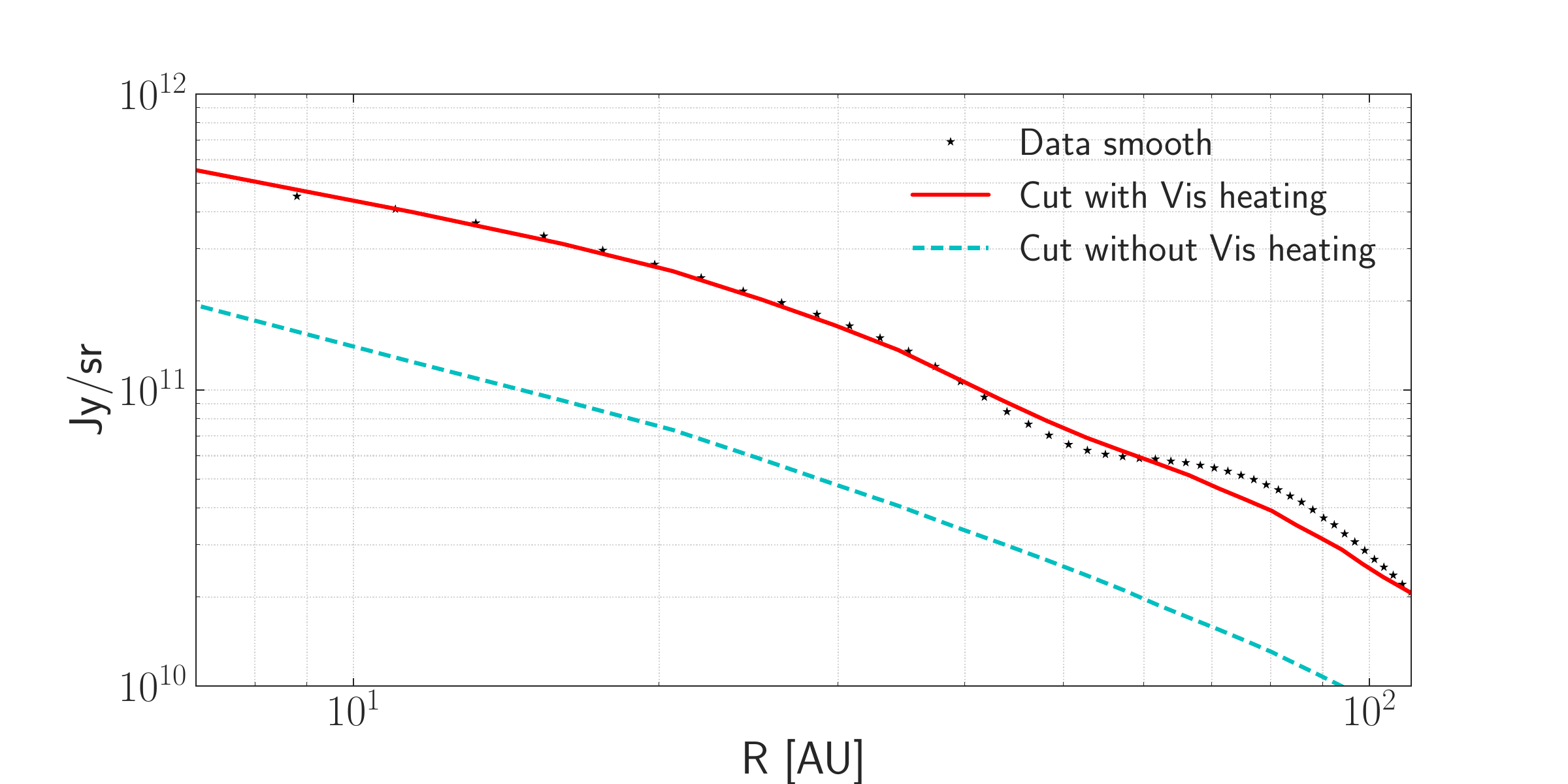}
\end{center}
\caption{Comparison between the radial profiles of the dust continuum emission at $\lambda=1.3$  \ mm and our models. We observe that our model with viscous heating is the best match for the observations while the passive heating is not enough to reach the observed emission.}
\label{fig:Diff}
\end{figure}

The synthetic predictions of the continuum images ( $\lambda=1.3$ mm ) in Figures \ref{fig: Images}  and \ref{fig:Diff} match the observations. The Figures illustrate that viscous heating causes a considerable change in dust emission when it is compared to the passive heating model. The viscous heating produces a centered increment in the intensity that decays at outer radii. The effects of viscous heating translate into a steeper emission radial profile, such as the one observed in V883 Ori.

The MCMC fit done by \citet{AESV} considers a model with a constant flaring index and without the addition of viscous heating. In general, the omission of the viscous heating for sources with typical accretion rates $\dot{M}$<10$^{-6}$ M$_{\odot}$ yr$^{-1}$ will not produce considerable changes in the thermal structure beyond $\approx$5 au. However, for rapidly accreting objects like FUors, in particular for V883 Ori, viscous heating will strongly affect the disc shape, their thermal structure, and therefore the dust continuum emission.

Overall, we reproduce the inner disc emission. However, we do not explain the wiggle at the border of the snowline. The wiggle could be produced by changes in the dust density or different dust populations due to dust evolution. Wiggles and phenomena like these have been predicted before in the literature  \citep{SnowlineBanzatti2015ApJ, Duts_Sintering_2016ApJ_Oku, COCO2,COCO1}. Those studies report the presence of density wiggles and dips before and after the snowlines, finding a greater concentration of dust just outside of them, and a small depletion on the inside. Such radial density fluctuations could then explain the observed wiggle in the dust emission.  However, understanding the specific changes in the dust composition around the snowline is beyond the scope of this paper.

Our model also has a sharp increment of the optical depth at r$\sim 40$ au. Besides a possible change in the opacity regime caused by the melting of the ice grains, another possible treatment for this issue could be a differentiation in the dust species. \citet{Notsu..22} show that there is a chemical enrichment of the ices behind the water snowline due to the recondensation of many volatiles, changing their composition. The change in dust population after the snowline produces a local change in the dust opacity. A thinner disc inside the snowline locally decreases its emission and heats up the dust right beyond the snowline by increasing its exposure to stellar irradiation. Thus, the intensity wiggles at 1.3 mm can also be explained by a dust population composed of larger ice grains beyond the snowline.

\section{Discussion}\label{Sect4}

From the analysis of molecular line emission, \citet{Merel, Tobin..Nature} have placed the location of the snowline between 80 and 100 au, while \citet{Snowline} estimated it $\sim$ 40 au from the dust continuum emission. The discrepancies in the location of the water snowline might be related in part to distinct scale heights traced by different observations together with the complex thermal structure of an outbursting source such as V883 Ori as shown by the results of our modeling.

We have shown that V883 Ori, as a FUor, has significant changes in its intrinsic structure compared to passively heated discs. The disc shape, accretion rates,  and the shift of the snowline make FUors particularly interesting for the study of their chemical and physical processes \citep{Molyarova}. Changes in the shape and the additional heating sources have significant effects on the disc emission even beyond the main region of influence. Figure \ref{fig:Diff} shows that viscous heating also increases the millimeter emission beyond 100 au. When viscous heating is not included, matching the brightness temperature causes the disc to be severely flared in the inner disc to match the emission. This compensation produces an important shadowing effect toward the outer disc making it cooler. In addition, a more extended hotter inner disc diffuses heat toward the outer disc, heating it further.

The V883 Ori disc structure and physical properties make it a suitable source for the study of the composition of the material that takes part in the planet formation process and the effects caused by disc evolution. For example, \citet{Merel} and \citet{LEE_V883} were able to characterize the detection of organic molecules far out in the disc, and \citet{Dary..et..al..2022} did a comprehensive study of the emission of multiple chemical tracers in V883 Ori, while \citet{Tobin..Nature} resolved the deuterium hydrogen oxide(HDO) emission coming from the disc. Multi-wavelength studies have the potential to shed light on the effect that the snowline has on dust populations and their possible role in the formation of pebbles and planet formation.

Our results show a variable accretion rate with radius. The recovered accretion ranges between different values, from by using SED fitting \citep{Liu..2022}, which are slightly lower than $\dot{M}_* \sim 7 \times 10^{-5}$ M$_{\odot}$ yr$^{-1}$ using the bolometric luminosity to  $\sim 1.1 \times 10^{-4}$  M$_{\odot}$ yr$^{-1}$  the peak of our retrieved accretion rate profile. A variable accretion rate with a radius least to pile-ups of material in the local minima of the radial profile. Variable accretion rates, hence, pile-ups of material are still consistent with the characteristic accretion outburst behavior of FUors. Pile-ups can trigger the MRI and the Thermal Instability. Moreover, theoretical  models predict that the outbursts are highly variable and episodic in nature \citep{NSAcc, Zhu..2010, Bae2013ApJ, HartmannIII}.

Even though viscous heating explains the steep radial trend in emission at the inner regions of the disc, the continuum emission has a wiggle at the boundary of the snowline. Changes in the composition and size distribution of the dust populations could be a possible solution to such variation in the radial emission profile. The sublimation of the icy mantles on dust grains makes dust growth less efficient, favoring the formation of smaller grains inside the snowline. Therefore, we expect changes in the dust opacity due to different dust populations or modulations of the surface density in and out of the snowline, producing the wiggle in the dust continuum emission. Water ice can increase dust stickiness, enhancing dust growth and changing the dust size distribution in the disc. The thermal inversion can produce distinct footprints in molecular absorption in infrared wavelengths in the inner disc\citep{Calvet..92} or in the brightness radial profiles of different ALMA continuum bands for example.

One of the caveats in our modeling approach is that we assumed the vertical density distribution of a passive disc. However, the vertical distribution may not follow a Gaussian distribution. \citet{Uzdensky..2013} described the vertical distribution of an active disc powered by the Magneto-Rotational Instability(MRI). In their assumptions, assuming hydrostatic equilibrium, the distribution of the active disc is proportional to $\propto (1 - \frac{z^2}{z_c^2})^3$. Even though the assumed prescription for the gas distribution will produce changes in the vertical temperature as well, the midplane temperature will only have changes within the same order of magnitude, while preserving a diffusive approach. Changing the heating input uniformly or following the gas density distribution did not make significant changes to the millimeter brightness temperature of the disc or the overall reported trends in this work, although it will impact other disc parameters locally.

Another source of uncertainty in our work is the distance to V883 Ori. The Gaia catalogs flag the astrometric solution of V883 as unreliable. Thus, we adopted the distance of 388 pc from \citet{Connelley_Reipurth, Kounkel..et..al..2017}, also consistent with the value derived by \citet{LEE_V883} from neighboring sources. Despite the points mentioned above, using the Gaia distance to V883 Ori (d$\sim$ 270 pc) does not qualitatively change our results. Such change would have mainly affected the values of the disc and stellar parameters, such as their masses, sizes, and the location of the snowline. If we have used Gaia's distance as the distance to V883 Ori, the snowline would have been at $\sim 26$ au, still farther than expected values. Thus, the need for an extra heat source besides passive heating still remains.

\section{Summary}\label{Sect5}

In this work, we report that when viscous heating is included in V883 Ori, the midplane temperature in the optically thick region is higher than the surface temperature. In the first few au, viscous heating increases the temperature to the order of 1000 K and higher values. In fact, viscous heating is the principal heating source at the inner disc, more important than the stellar or flaring terms. However, for stars with lower accretion rates ($\dot{M}<10^{-6}$ M$_{\odot}$ yr$^{-1}$), the stellar heating will still be the main heating source in most parts of the disc (see Fig. \ref{fig:V_vs_S}). 

The results of our semi-analytical approach show a strong initial flaring, and then a plateau of the aspect ratio of the disc in the inner 30 au. After the initial 30 au, the disc starts to self-shadow. In that region, the disc has a flaring index lower than one, i.e., the aspect ratio diminishes with radius, as it is shown in Fig. \ref{fig: Self-Consistent}. We conclude that viscous heating plays a significant role in shaping the disc structure.

 We compare the radiative transfer predictions of the model with the observations in Figure \ref{fig:Diff}. From such comparisons we can conclude that viscous heating provides the extra energy budget to reach the emission levels, showing that the addition of viscous heat increments the continuum emission as a central source close to the star by changing the radial slope of the thermal profile, making it steeper.

In summary, classical passive heat sources are not capable of reproducing the emission profile of V883 Ori, an additional heating source is needed. This extra source has the particularity of being significant in the inner part, centrally peak and it decays quickly after a few au. High accretion rates are typical of FUors and their associated viscous heating matches the necessary conditions to explain the dust continuum emission profile of V883 Ori. Viscous heating not only heats up the disc but also changes its physical structure. V883 Ori in particular also presents a wiggle right at the snowline boundary where dust evolution could play a major role in creating dust substructures.

\section*{Acknowledgements}
S.C. acknowledges support from Agencia Nacional de Investigaci\'on y Desarrollo de Chile
(ANID) given by FONDECYT Regular grant 1211496, ANID project Data
Observatory Foundation DO210001. S.C., S.P. and L.C. acknowledge support from Millennium Science Initiative Program -- Center Code NCN2021\_080.
NCN2021 080. W.L. acknowledges support from the NASA Theoretical and Computational Astrophysical Networks (TCAN) via grant 80NSSC21K0497, from the NASA Emerging Worlds program via grant 22-EW22-0005, and by NSF via grant AST-2007422. S.P. acknowledges support from FONDECYT Regular grant 1231663 and ANID. L.C. acknowledges support from FONDECYT  Regular grant 1211656. The authors also acknowledge useful discussions with Luca Ricci and Neal Turner. 

This paper makes use of the following ALMA data: ADS/JAO.ALMA\#2013.1.00710.S,  ADS/JAO.ALMA\#2015.1.00350.S. ALMA is a partnership of ESO (representing its member states), NSF (USA) and NINS (Japan), together with NRC (Canada), MOST and ASIAA (Taiwan), and KASI (Republic of Korea), in cooperation with the Republic of Chile. The Joint ALMA Observatory is operated by ESO, AUI/NRAO and NAOJ.

\section*{Data Availability}

The results and outcomes of the models can be reproduced by the calculations explained in the article. The observational data is public and available in the ALMA Archive at \href{https://almascience.nrao.edu/aq/}{https://almascience.nrao.edu/aq/}.  The product files of the models are also available upon reasonable request to the corresponding author.

\bibliographystyle{mnras}
\bibliography{ref}

\appendix

\section{Fitting Algorithm Optimization}\label{App}

\subsection{Self-Consistent Algorithm}\label{App:1}

Initially, we use a self-consistent two layers model as in \citet{ChiangSED1997ApJ} as the starting point to find the thermal structure in the disc. We consider the disc as being passively heated, i.e., without viscous heating. We evaluate the irradiation heating with a stellar radius of $R_* = 4R_{\odot}$ and a stellar temperature of $T_* = 7000 \ K$. Equation \eqref{stel_plus_flar} takes into consideration the effect of the stellar size and the disc shape. The $\Big(\frac{dh}{dr} -\frac{h}{r} \Big)$ term states  that the flaring could actually cool down the disc when it is self-shadowed, i.e.,  $\frac{dh}{dr} < \frac{h}{r}$, which may occur in V883 Ori. Additionally, the flaring term shows that the irradiation heating is dependent on the local scale height and its radial gradient. Once we have calculated a midplane temperature, the sound speed changes, which leads to an update of the scale height and  the flaring. The thermal dependence of the dust opacity is also considered in the algorithm.

 The self-iterative algorithm starts by solving the effective temperature for each radius from the outer part to the inner regions. It starts fitting with two layers once the disc gets optically thick at mm wavelengths, i.e., $\tau_{\rm mm}> 1$.  For the first run, we only consider irradiation and basal terms. After that, we add the viscous heating term by equating the brightness temperature to the accretional temperature in the regions where $\tau_{\rm mm}>1$. Then, the midplane temperature is calculated using Equation \eqref{eq:: T_mid} where the Rosseland mean opacity is calculated following the \citet{Semenov..2003} opacities. The dust density is taken from the fit done by \citet{AESV}. Changing the temperature produces a change in the scale height and the sound speed, which in turn changes the flaring and aspect ratio, so the temperature varies again. The self-iterative algorithm takes into account the changes in the local flaring, aspect ratio, and opacity during each iteration and updates them with the new calculated temperature. After an iteration is done, the algorithm starts again from the beginning with a new irradiation term  that is given by the new shape of the disc (see Eq. \eqref{stel_plus_flar}). The algorithm follows iterating until the thermal difference between iterations is smaller than 0.1 K.

Reaching convergence in the cold outer disc becomes harder because the heating terms are smaller, and changes in flaring and aspect ratio are less abrupt. Before running the algorithm, we study its convergence considering different initial outer radii.  We found that the algorithm converges in the inner $\approx 150$ au,  so using $r_{\rm out}$=500 au as the outer boundary for the grid is good enough to trace the temperature inside the snowline. The algorithm uses a background temperature of $T_b=10$ K at the outer radius, which acts as a lower bound for the disc temperature. We try different temperatures for this background temperature between $T=10$ K and $T=20$ K without significant changes in the final result. For the initial runs, we set the algorithm to start with a self-iterative loop from the outer (500 au) to the inner disc (outside-in) with an initial flaring index of 1.5 and an initial aspect ratio of 0.13.

\begin{figure}
	\centering
 \includegraphics[height=1.4\linewidth]{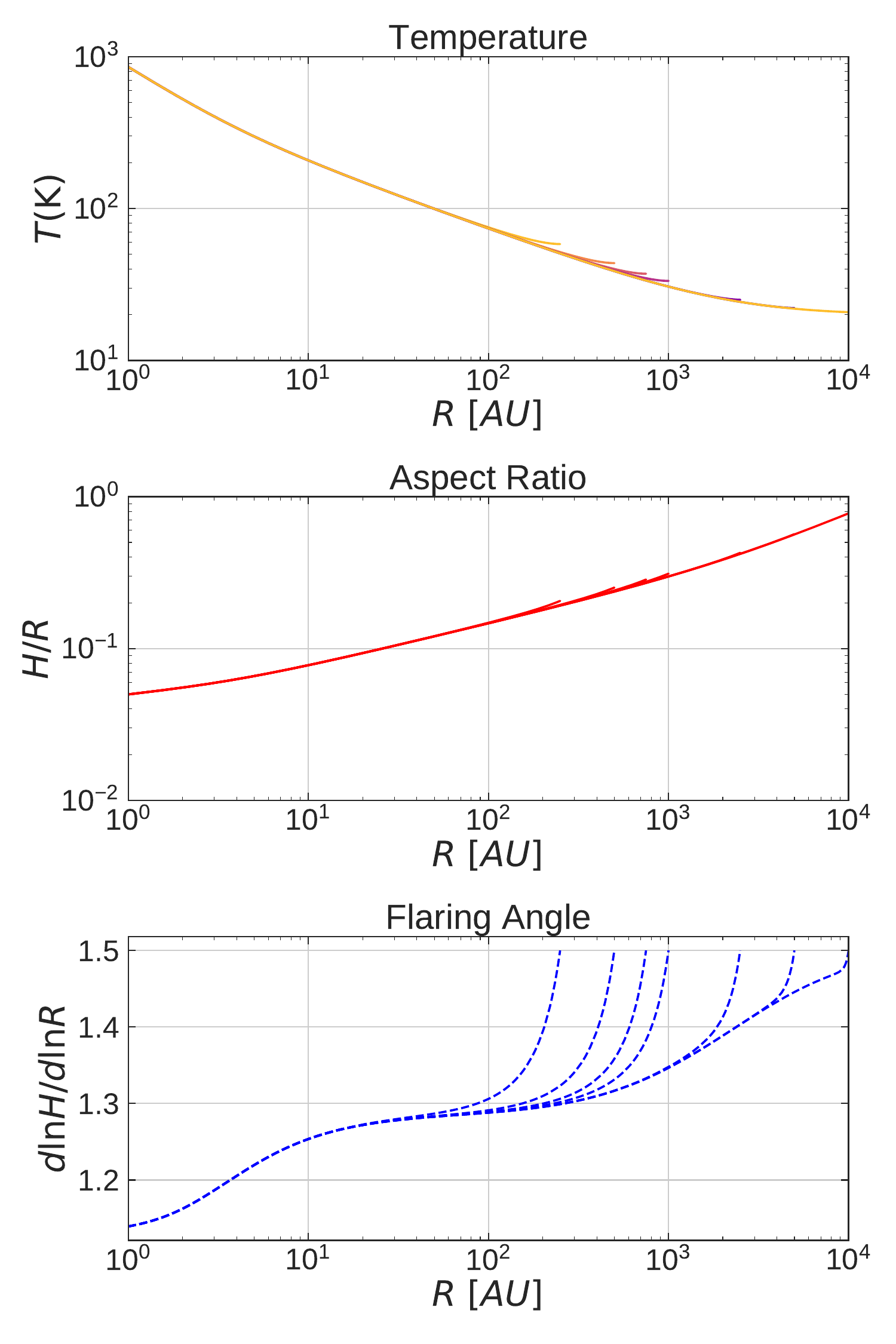}

\caption{Study of the convergence of the algorithm for different outer radii. \textbf{Top:} Temperature, \textbf{Middle:} Aspect Ratio, \textbf{Bottom:} Flaring Index. The aspect ratio and temperature converge quickly no matter the initial outer radius given for the algorithm. The flaring index is more sensitive to the starting point of the algorithm, so we set the initial outer radius at $r_{\rm out}=180$ au. By doing so we secure reaching convergence inside the snowline.}
\label{fig: Convergence}
\end{figure}

To test the convergence if the disc shape in the model, we run the algorithm with different outer radii as starting points, ranging from 110 to 10000 au. Figure \ref{fig: Convergence} shows the convergence of the algorithm for a sample of different outer radii in that range. We note that for the temperature and the aspect ratio, the algorithm converges very quickly to a stable solution. Inside 100 au, all of them have already converged. The flaring index takes the longest to converge and it is also the most variable, so we use this parameter to define the radial extension for the fit. Regarding the flaring, our model starts with a flaring index of 1.5 at the outer radius of the grid and then converges to a value of 1.15. Considering the behavior of the algorithm shown in Figure \ref{fig: Convergence}, and given that we are interested principally in the modeling of the inner 50 au, the convergence of the algorithm is not an important concern for the modeling.

\subsection{Bootstrapping}\label{App:2}

We also fitted the emission with a bootstrapping algorithm by taking 1000 samples from gaussian distributions  centered at the fitted parameters (Temperature and surface density) obtained from a self-iterative algorithm for the midplane temperature and the surface density, assuming a 20\% deviation from those values. We then proceed to do the fitting at all radii by minimizing the $\chi^2$ metric from the observed and predicted brightness temperature i.e.:

\begin{equation}
   \chi^2 = \frac{1}{N}\sum (T_{\mathrm{b,data}}-T_{\mathrm{b,predicted}})^2,
\end{equation}

\noindent where $T_{\mathrm{b,data}}$ is the brightness temperature from the 1.3 mm continuum data, $T_{\mathrm{b,predicted}}$ is the temperature predicted by the fitting, and $N$ the number of sampled radial points. In the bootstrapping, all points are fitted simultaneously and it including passive  and active heating sources together with radial and vertical diffusion.

\section{Disk structure without viscous heating}\label{App:3}

\begin{figure}
    \centering
    \includegraphics[width=0.99\linewidth]{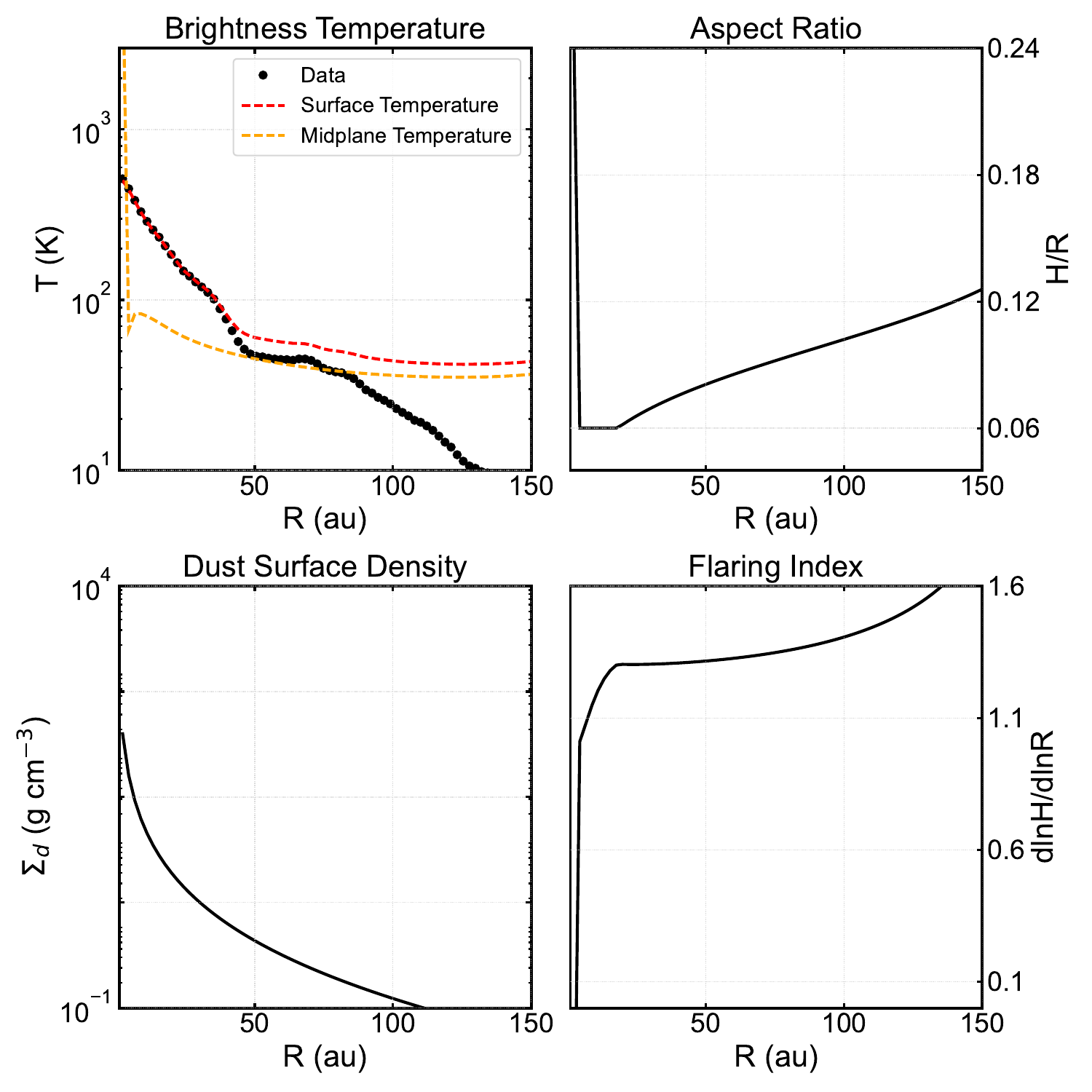}
    \caption{Similar to Figure \ref{fig: Self-Consistent} but without viscous heating included. The mass accretion rate and infall panels are excluded since they are not part of the fitting when viscous heating is not considered.}
    \label{fig:No_vh}
\end{figure}

We show the structure of the disk when viscous heating is not included in the fitting in Figure \ref{fig:No_vh}. A similar approach was already addressed by \citet{AESV} showing that the fitting of a parametric disk compensates for the inner emission by an overall self-shadowing disk. In our fit we observed that in order to match the inner disk emission, the disk would have an inner wall with a very high aspect ratio in the inner 10 au, which would shield the outer disk making it cooler than the counterpart with viscous heating.

\end{document}